\documentclass[preprint,showpacs,preprintnumbers,amsmath,amssymb]{revtex4}

\usepackage{graphicx}
\usepackage{dcolumn}
\usepackage{bm}

\begin{document}
\title{Total electronic energy by tight binding approximation and experimental
toughness  of three different hybrid polymers}

\author{   N.Olivi-Tran$^{1,2}$,A.Ferchichi$^2$,S.Calas$^2$,P.Etienne$^2$}
\affiliation{$^1$S.P.C.T.S., UMR-CNRS 6638, Universite de Limoges, 47 avenue Albert Thomas
87065 Limoges cedex, France}
\affiliation{$^2$G.E.S., UMR-CNRS 5650, Universite Montpellier II, 
case courrier 074, place Eugene Bataillon, 34095 Montpellier cedex 05, France
}
\date{\today}

\begin{abstract}
We computed by a modified tight binding approximation, the total electronic
energy of three different hybrid polymers: $H-SiO_2$, $CH_3-SiO_2$ and
$C_6H_5-SiO_2$. We made the hypothesis that the structures of these polymers
are amorphous. Computational results regarding the total electronic
energy and experimental data \cite{ferchichi} on the toughness of these
three hybrid polymers were compared. A good qualitative agreement
was found between computations and experiments.
\end{abstract}
\maketitle
Since few years, organic inorganic hybrid materials issued from sol-gel process and an organic polymerized part are more and more intensively studied.
They offer a very innovative way to develop a wide variety of new materials
because of their structure at the nanometer scale which combines
the properties of an organic and an inorganic entity.

On an industrial point of view, these materials are more used as layers such as protective coatings \cite{1}, materials with high transparency \cite{2}
transistors \cite{3}, luminescent diodes \cite{4}, solar cells \cite{5},
waveguides \cite{6} and photochromic coatings \cite{7}.

Hybrid materials may be classified into two families. The class I family
corresponds to hybrid materials where the organic part is embedded
in an inorganic network. The interactions between the mineral and the organic
parts are weak essentially Van der Waals, hydrogen bonds and electrostatic
interactions \cite{43}.
The class II corresponds to the existence of chemical bonds (covalent or ionic-covalent) between the organic and the mineral part of the network \cite{43}.
The synthesis of class II hybrid polymers has been initiated simultaneously
by the sol gel scientists and the polymer scientists.

The sol gel process is a method to obtain hybrid polymers: one would have
to incorporate to the sol inorganic precursors and organic compounds
with functionalities which can be plugged to the inorganic part
of the gel. This may lead to hybrid nanomaterials \cite{39}.

We deal here with class II hybrid polymers containing the species: $Si,C,O$ and $H$. This type of polymers may be obtained by the sol gel process.
In this particular type of polymers containing only $sp^2$ and $sp^3$ bonds it is possible
to use a tight binding approach to compute the total electronic energy.

Our tight binding method has been modified in order to take into account
hybridization i.e. the $\sigma$ and the $\pi$ valence electrons which
enter a covalent bond. The tight binding method depends only on the connectivity
of the atoms which enter a structure and not on the real distribution
of the atoms in space. In a previous work, we showed that the connectivity
of hybrid polymers containing only $Si,C,H$ and $O$ was amorphous \cite{olivitran}.

The aim of this study is to compare the computed toughness of three kinds
of hybrid polymers to the toughness of the same kind of materials
obtained in the literature.

Let us introduce our tight binding approach.
Let us remind that this is a one electron model, each electron moves in a
 mean potential  $V(r)$ which represents both the nuclei attraction and 
the repulsion of other electrons. $\sigma$ and $\pi$ electrons are separately treated :

If the molecular orbital $\sigma$ is given by:
\begin{equation}
|\Psi >= \sum_{i,J} a_{iJ} | iJ >
\end{equation}

and the energy origin taken at the vacuum level, the Hamiltonian can be written as, in the case of $sp^{\nu}$ ($\nu = 1,2,3$) hybridization :
\begin{equation}
H_{\sigma} = E_m \sum_{i,J} | iJ >< iJ |  +  \Delta_i{\sigma}\sum_{i,J,J' \neq J} | iJ >< iJ | + \beta_{\sigma}\sum_{i,i'\neq i,J}  | iJ >< iJ |
\end{equation}
($i$ and $i'$ are first neighbours) where $E_m$ is the average energy: 
$ E_m =  (E_s - \nu E_p) / (\nu + 1)$ ,$E_s$ and $E_p$ are the atomic level energies,$\beta_{\sigma}$  is the usual hopping or resonance integral in tight binding theory (interaction between nearest neighbour atoms along the bond),
$\Delta_s$ is a promotion integral (transfer between hybrid orbitals
 on the same site) :$\Delta_{\sigma} =  (E_s - E_p) / (\nu + 1)$.

The Hamiltonian of the $\pi$  bonds is given by:
\begin{equation}
H_{\pi} = E_p \sum_i  | i >< i |  + \beta_{\pi}\sum_{i,i'\neq i}  | i >< i'|
\end{equation}
with $ | i >$  the $\pi$  orbital centered on atom $i$,and $\beta_{\pi}$ the hopping integral for $\pi$ levels.

 We need only 3 parameters: $\beta_{\sigma}$, $\beta_{\pi}$, and $\Delta_{\sigma}$
  for the homonuclear model which represent in fact the average potential $V(r)$ and which take into account the nuclear attraction and the dielectronic
 interactions \cite{joyes}. But due to the fact that we only take into account
 on average the nuclear and dielectronic interactions, we can only compare
 clusters with the same number of atoms.

The numerical values of the parameters are given in table 1.

In the following are the computational results compared to experiments.
We computed the total electronic energy for three types of hybrid polymers:
$H-Si$, $CH_3-Si$ and $C_6H_5-Si$ for the organic part of the hybrid polymers
and $Si-O-Si$ for the inorganic part of the hybrid polymer.

In figure 1, one may see the typical structure that we used for the tight
binding calculation in the case of an amorphous hybrid polymer.
The picture shows a planar molecule but this may be folded and the angles
between different atoms may not be equal to 90$^o$ and the length of the bonds 
may be changed depending on the type of atoms \cite{86,87,88,96}. Thus it represents
an amorphous structure. Here $R=H,CH_3$ or $C_6H_5$. Let us remark that there 
are $sp^2$ bonds in the  $C_6H_5$ cycle.

In figure 2, we showed the total  electronic energy as a function of the number
of atoms in the polymer.

Figure 3 shows again the total electronic energy but as a function
of the number of valence electrons  (not differentiating the $sp^2$
and $sp^3$ bonds i.e. the $\pi$ and the $\sigma$ electrons).

Finally, table 2 is a comparison of the toughness of the three different
hybrid polymers that we computed here, with the experimental data
coming from the literature \cite{ferchichi}.

We made a linear regression of the results shown in figure 3. The result
is that the slope of the total electronic energy as a function
of valence electrons is the same within computational error
for $H-Si$ and $CH_3-Si$: $22,0\pm 0.6 eV$. Thus the difference
of total electronic energy cannot be differentiated by the the 
number of valence electrons. But, in the case of $C_6H_5-Si$,
the slope given by the linear regression is $20.9\pm0.6 eV$.
We can conclude from figure 3 that the total electronic energy
is larger for the case $C_6H_5-Si$ than for the two other cases
of hybrid polymers.

This result may be related to mechanical properties of such material:
the toughness of the amorphous hybrid material is the largest; indeed,
the total electronic energy is related to the toughness of the electronic
bonds within the structure, so the toughness of the electronic bonds
can be linked to the mechanical toughness of the material.

Let us examine more in details the total electronic energy as a function
of the total number of atoms. We made once again a linear regression
over the three different hybrid polymers.
For $H-Si$, we obtained:
\begin{equation}
E=217,4-89.7 .N_{at}
\end{equation}
For $CH_3-Si$ the result is:
\begin{equation}
E=382,5-79,6.N_{at}
\end{equation}
and finally for $C_6H_5-Si$, we obtained:
\begin{equation}
E=368,5-72,3.N_{at}
\end{equation}
where $E$ is the total electronic energy in $eV$ and $N_{at}$ is the total
number of atoms.
We see that in the case of $C_6H_5-Si$, the slope is the largest (do not forget
the minus sign), therefore, the stability of such hybrid polymer is the smallest
compared to the two others. This is in good agreement with the linear regression
of figure 3 given before.
But, in figure 3 one cannot distinguish the stability of $H-Si$ and $CH_3-Si$.
With the results of figure 2, this is done: $CH_3-Si$ is less stable than
$H-Si$ as its total electronic energy is larger for a large number of atoms.

We compared the toughness of our computed hybrid polymers with the results
given by literature \cite{ferchichi}.
The results are given in table 2. 
As one may see in table 2, the total electronic energy and the experimental
toughness follow the same tendency. The toughness of $H-Si$ and $CH_3-Si$
are in good agreement with the computed total electronic energy.
Regarding the results of $C_6H_5-Si$, one cannot make a direct
comparison: the comparison cannot be quantitative.
For that we calculated the ratio $E/T$ and we see that for the cases
of $H-Si$ and $CH_3-Si$, the ratio is almost the same, while for
the case of $C_6H_5-Si$ it is three times larger.

Indeed, the total electronic energy of $C_6H_5-Si$ is the largest (thus
the less stable) but the numerical results compared to the toughness
can only be compared qualitatively.
The explanation of this feature is that experimentally \cite{ferchichi}
the presence of the $C_6H_5$ group leads to a less connected array
in the final polymer obtained by sol gel \cite{ferchichi}.
As we did not take into account this type of phenomenon (the connectivity
was an hypothesis of our numerical computation), we did only compute
the total electronic energy of an ideal structure.
In real systems obtained by sol gel, 20\% of the species are not totally
condensed in the case of $C_6H_5-Si$ and 5\% to 10\% are not condensed
for $H-Si$ and $CH_3-Si$ \cite{ferchichi}.
This explains the difference between computation and experiments.

To conclude, we may say that the toughness of experimental and numerical
such hybrid polymers may be compared qualitatively: the toughness
(and $E$) of $H-Si$ is the largest followed by $CH_3-Si$ and finally
the smallest toughness if obtained for $C_6H_5-Si$.

A final conclusion is that we
 made a comparison between the toughness of three different hybrid
polymers and the total electronic energy obtained by a modified
tight binding method.
The result is that the toughness follows the same tendency
as $E$ but only qualitatively. This is due to experimental
characteristics of the sol gel process which is used to obtain
these polymers  which we do not take into account in our computations.

Acknowledgements: N.O.T. would like to acknowledge M.Leleyter for helpful discussions

\pagebreak
\begin{table}
\begin{center}
\begin{tabular}{|c|c|c|c|c|}
\hline
 atom/parameter & $E_{\sigma}$ & $E_{\pi}$ & $\beta_{\sigma}$ & $\beta_{\pi}$ \\
\hline
$H$ & 13.6 & 0.0 & 15.05 & 0.0 \\
\hline
$C$ & 19.45 & 10.74 & 7.03 & 3.07 \\
\hline
$O$ & 32.37 & 14.96 & 12.0 & 5.0 \\
\hline
$Si$ & 14.96 & 7.75 & 4.17 & 0.8 \\
\hline 
\end{tabular}
\end{center}
\caption{Parameters for the tight binding calculations}
\end{table}
\begin{table}
\begin{center}
\begin{tabular}{|c|c|c|c|}
\hline
hybrid polymer & total electronic energy per mol $H$ ($J$) & experimental toughness $T$ ($MPa.m^{1/2}$) \cite{ferchichi} & ratio $H/T$ \\
\hline
$H-Si$ & $-3.3.10^6$ & $0.33\pm 0.05$ & $-10^7$ \\
\hline
$CH_3-Si$ & $-2.98.10^6$ & $0.32 \pm 0.05 $ & $-0.93.10^7$  \\
\hline
$C_6H_5-Si$ & $-2.71.10^6$ & $0.09\pm0.01$ & $-3.10^7$ \\
\hline
\end{tabular}
\end{center}
\caption{Comparison between the total electronic energy of the three hybrid
polymers computed by tight binding and experimental toughness from literature \cite{ferchichi}}
\end{table}
\begin{figure}
\includegraphics[width=18cm]{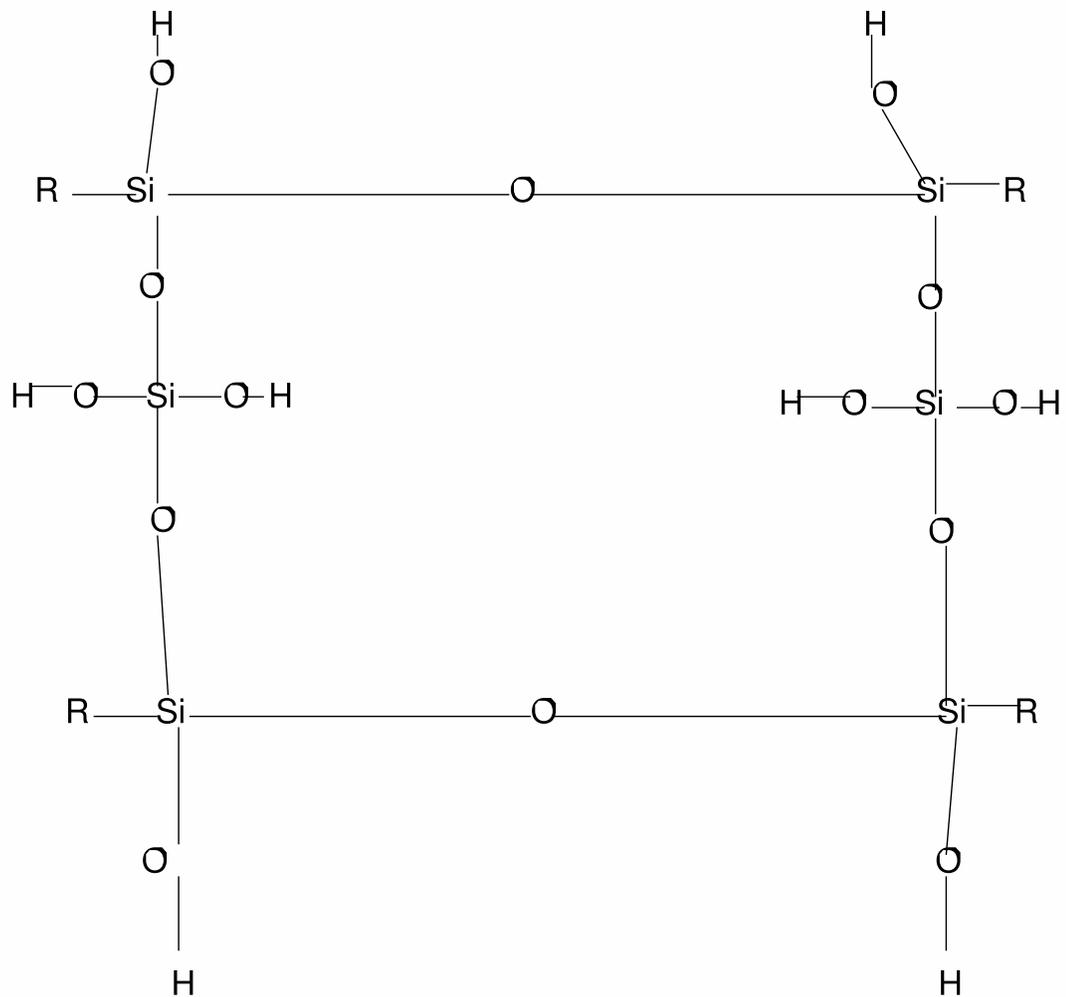}
\caption{Example of amorphous molecule with $R=H,CH_3,C_6H_5$}
\end{figure}
\begin{figure}
\includegraphics[width=18cm]{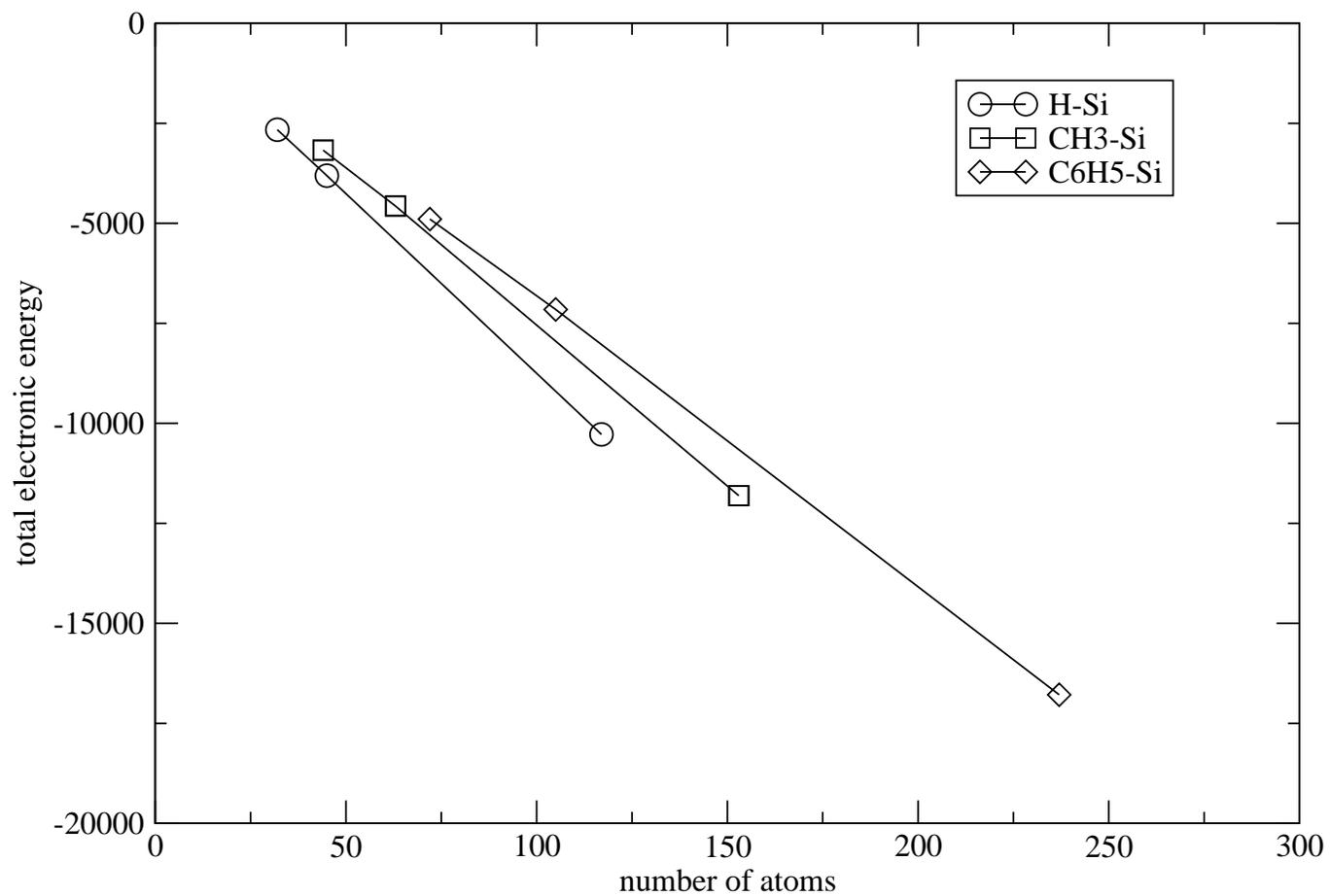}
\caption{Total electronic energy as a function of the number of atoms}
\end{figure}
\begin{figure}
\includegraphics[width=18cm]{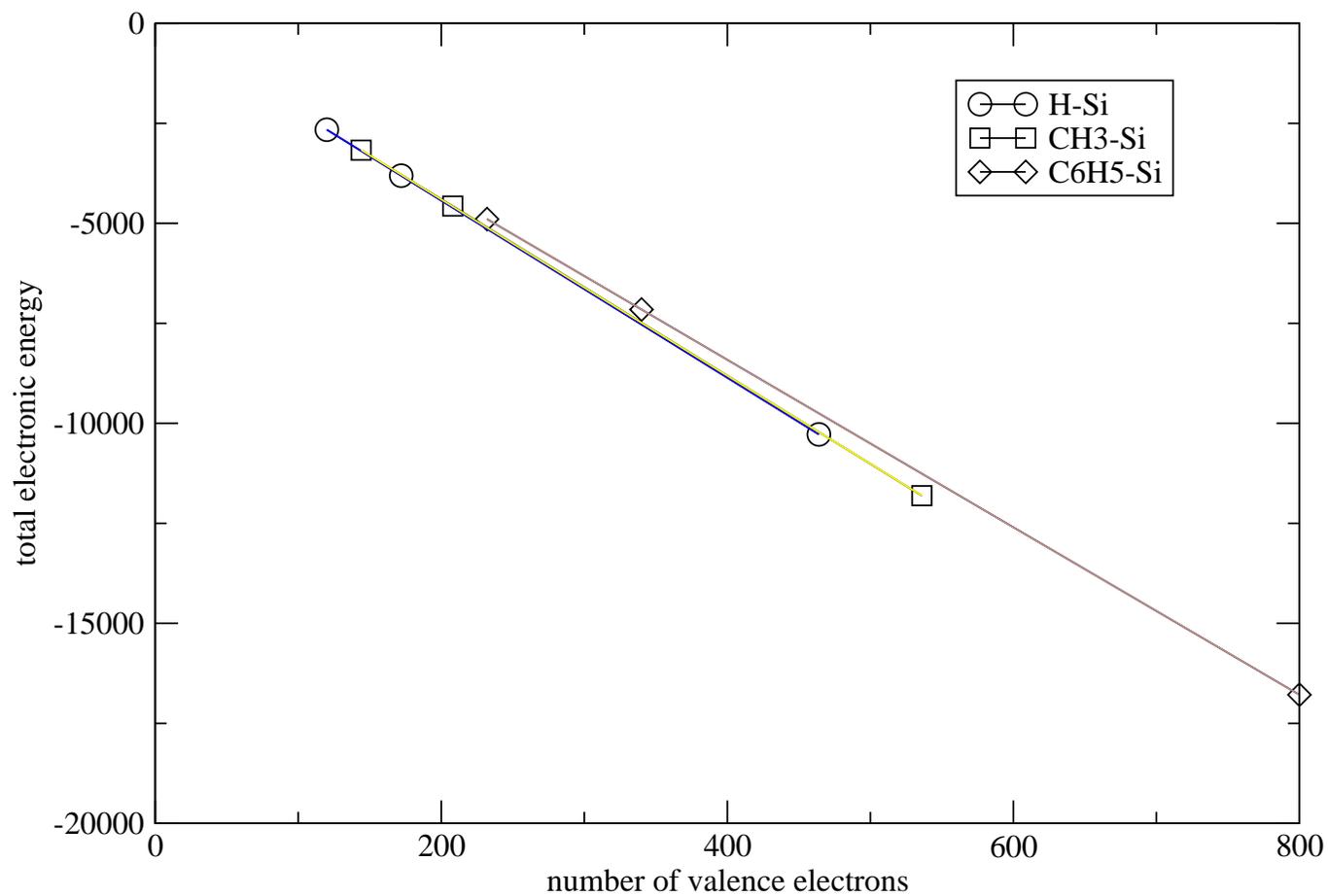}
\caption{Total electronic energy as a function of the number of bonds ($\pi$ or
$\sigma$)}
\end{figure}

\end{document}